\documentclass[12pt,preprint]{aastex}

\shorttitle{A tiny Galactic dust cloud projected onto NGC 3269?}
\shortauthors{Dirsch et al.}

\begin{document}

\title{A tiny Galactic dust cloud projected onto NGC\,3269?}

\author{B. Dirsch, T. Richtler and M. G\'omez}
\affil{Astronomy Group, Depto. de F\'{\i}sica, Universidad de Concepci\'on,\\
Casilla 160-C, Concepci\'on, Chile}

\begin{abstract}
We report on new observations obtained with the Magellan Clay-telescope of a
tiny dust patch in the Sa-galaxy NGC 3269 in the Antlia galaxy cluster.
It has already been suspected to be a projected Galactic foreground cloud. 
In this case a size of about 4 arcsec and a maximum absorption of 
$\sim$1 mag in the B-band would make it unique so far. We present 
further evidence for its Galactic nature from imaging under excellent 
seeing conditions (0.3-0.6 arcsec). This dust cloud could be the first 
optically identified counterpart of a new type of molecular cloud  
recently discovered by \cite{heithausen04}.
\end{abstract}


\section{Introduction}

In the course of studying the globular cluster systems of NGC 3258 and
NGC 3268 in the Antlia cluster, we noted a tiny dust patch in the neighbouring
 galaxy NGC 3269 \citep{dirsch03} with a diameter of about 4 arcsec, barely visible on 
 images found in the literature. 

This is striking since NGC 3269 is a grand design Sa spiral without signs
of star formation or small scale substructure. Moreover, the galaxy seems to be devoid
of HI \citep{barnes01}, who quote from their ATCA observations an upper limit of
about $\rm 10^9  M_{\odot}$. 
 We estimated the absorption in Washington C and Kron-Cousins R
and found a reddening law less grey than the standard one \citep{he95}.
The average absorption in R is about 0.3 mag, comparable to the average
foreground absorption in the line-of-sight.
At the distance of NGC 3269, this patch would
be a large dust cloud with a diameter of about 500 pc. Besides the fact that the existence of
a singular dust complex of this dimension is strange by itself, it should be accompanied
by star formation and should have many cores with very high line-of-sight
absorption, causing the average R-value to be larger, i.e. towards grey
absorption. Even more so, if the dust cloud were located inside the galaxy
with some stellar light in front of it.
These reasonings gave rise to  the conjecture that this cloud could be a Galactic
one in the foreground, projected onto NGC 3269. The low galactic latitude
of NGC 3269 (-19$^{\circ}$) supports this interpretation. To our knowledge this would
be the first detection of an isolated  Galactic dust cloud of this angular
size (100 pc distance: 4 $\arcsec$ equal 400 AU). A larger dust complex in this region is
not obvious. In the catalog of Galactic dust clouds by \cite{dutra02} the next entry near to NGC 3269 is the IR-excess cloud IREC403 \citep{desert88} whose center is 78 arcmin away.
IRAS maps show 100$\mu$-emission about 5$\arcmin$ north of NGC 3269 and about
10$\arcmin$ to the south-west. NGC 3269 itself is projected towards a local minimum of
emission.   

An accidental projection of a Galactic dust cloud onto a galaxy may by itself be
of no great interest. However, \cite{heithausen04} reported  the
discovery of tiny molecular clumps with densities indicating a high overpressure 
with respect to the ambient interstellar medium (ISM) and sizes resembling that 
of our
dust patch. Their properties suggest that they may be an abundant structural feature
of the ISM. Our dust patch might be the first optical identification
of this kind of molecular cloud. Therefore, one would like to have further support
or counterevidence for this hypothesis.
The seeing on our Washington images was about 1 arcsec and measurements on
scales of the order of the seeing may be unreliable.
In this contribution, we report
new observations under extremely good seeing to call further attention to this
apparently rare phenomenon.

\section{Observations}

During an observing run at the 6.5m Magellan Clay telescope at Las Campanas
 Observatories,
Chile, we took advantage
of a period of excellent seeing and imaged NGC 3269 in the Harris filters B and R with the MagIC camera.
Observing date was January 31, 2004. The CCD has 
a pixel scale  of  0.069 arcsec/pixel, providing a field of view of $2.36\arcmin\times2.36\arcmin$.
 The total exposure time
in B was 1800s, and 1200s in R.  
The frames were bias-subtracted and flatfield corrected using the 
IRAF
\footnote{IRAF is distributed by the National Optical Astronomy 
Observatories, which are operated by the Association of Universities for 
Research in Astronomy, Inc., under cooperative agreement with the National 
Science Foundation.}
 script MagIC-tools\footnote{
http://www.lco.cl/lco/magellan/instruments/MAGIC/iraf\_reductions.html}.
The seeing on the B-image was 0.35", on the R-image 0.6". Figure.\ref{image} shows the
dust patch in the south-western part in NGC 3269 in the B-band. In this image, the
sky has been subtracted. Further substructure
is visible. The shape is not spherical and the points of highest absorption are
located somewhat off-center.
The galaxy exhibits a very smooth stellar light
distribution which is distorted only by our
dust cloud and two extremely small dust cloudlets
($0.5\arcsec$) nearby (marked in Fig.\ref{image}). 
The inset shows NGC 3269 on a larger scale in Washington C (see \cite{dirsch03}
for details). Clearly visible are extended arc-like structures indicating 
previous or ongoing tidal interactions. These
events might have been responsible also for a gas removal out of NGC 3269.

\section{Reddening}

\begin{figure}
\centering
\includegraphics[width=14cm,angle=-90]{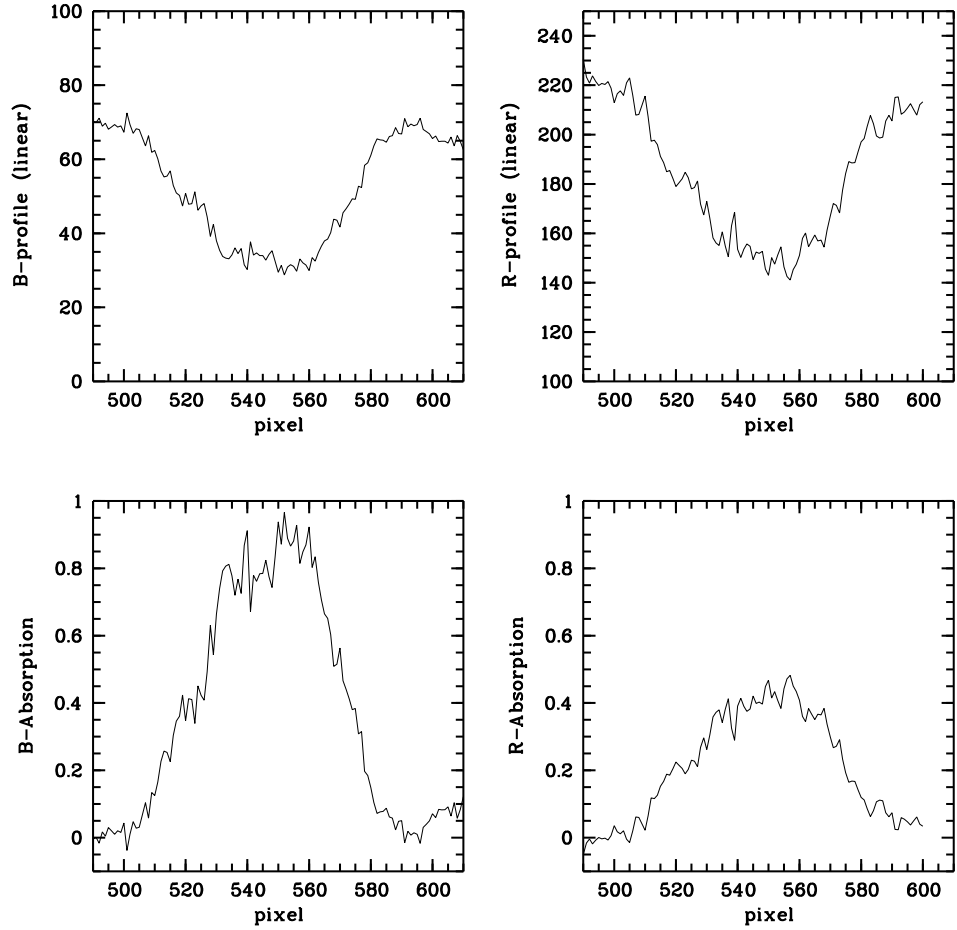} 
\caption{The upper panel shows the drop in intensity across the dust patch in
linear, arbitrary units,
averaged over 30 pixel (left: B-band, right: R-band). The lower
panel shows the relative absorption in mag. North is to the right,} 

\label{slit}
\end{figure}

The dust patch is not projected onto a homogeneously bright surface, which
would ease measuring the absorption. There is a gradient in surface brightness
in the East-West direction. It is smallest in the North-South direction which
we therefore chose to place a slit across the patch. The slit has a length of
210 pixels and a width of 30 pixels (corresponding to 14.5$\times$2.0 arcsec).
We take the average along the width.
Figure \ref{slit} displays the result.

The upper left graph shows the B-profile on a linear scale after sky subtraction.
To measure the relative absorption, we adopt a background level of 70.
The upper right panel shows the same for the R-image (note that 
the scale is now different). We adopt 220 as the local background value.
The lower panel shows the relative absorption in B and R.

\begin{figure}
\centering
\includegraphics[width=12cm, angle=00]{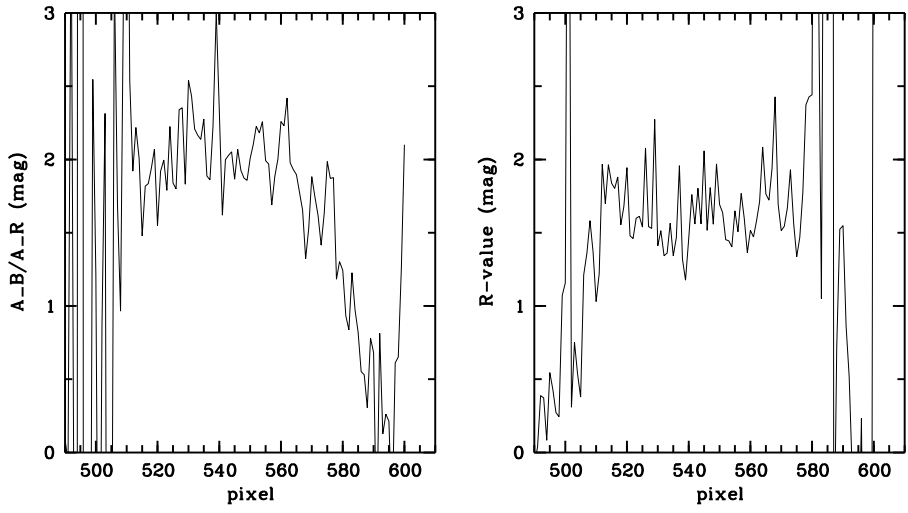}
\caption{The left graph shows the $A_B/A_R$-value, the right graph the 
R-value $\frac{A_B}{A_B-A_R}$. 
}

\label{absorption}
\end{figure}

With these values, we calculate the average absorption ratio $A_B/A_R$ and
the R-value $\frac{A_B}{A_B-A_R}$, which are shown in Fig.\ref{absorption} in
the left and the right graph, respectively. For $A_B/A_R$, the uncertainty on dividing
 two small numbers on the left side becomes evident. This is less distinct on the
 right side because here the boundary of the dust patch is relatively sharp. 
The worse seeing in R lets the R-absorption less rapidly
reaching zero than the B-absorption. We consider only the ''plateau'' in 
the pixel interval 520-570. In this region, $A_B/A_R$ is $2.00\pm0.28$ where 
the uncertainty is the standard
deviation. The uncertainty of this average value is difficult to formalize, because
the intensities are not independent, but should be much smaller than 0.25 mag.
Our Harris filters match Johnson B and Cousins R. A standard reddening law in these
filters is given by \cite{he95}. 
 The standard value
 of $A_B/A_R$  is 1.65, very similar to that given by \citet{rieke85} for the Johnson system. Given that we average
over small structure, which additionally makes the derived $A_B/A_R$-value greyer,
 the real value for $A_B/A_R$ will be higher (not lower!)
by an unknown amount. 

The right panel displays $R = \frac{A_B}{E(B-R)}$, and we measure
an average of $1.6\pm0.25$, again in the the pixel interval 520-570.
 The standard value according to \cite{he95} is 2.5, compared to 2.3 in the Johnson 
system \citep{rieke85}.

The statistical uncertainty alone would be small enough to conclude that the
reddening law is indeed less grey than the standard reddening law. 
The systematic uncertainty is dominated by the adoption of a sky value. 
The sky values, measured in the outer regions of the frames, are 176 in the
B-frame and 1160 in the R-frame.
The relative accuracy of the sky determination for the R-frame is
below 1\%, for the B-frame about 1\%. Adopting 1\% for both bands, translates 
into uncertainties of the absolute level of $\pm$10 units in the R-band and $\pm$2 units
in the B-band. Error propagation gives an uncertainty of 0.05 in both $A_R$ and
$A_B$, leading to about 0.1 for $A_B/A_R$ and $\frac{A_B}{E(B-R)}$.

Another factor is the local background. However, as can be seen from the 
scatter in the upper panels of Fig. \ref{slit}, any uncertainty is
not larger than the uncertainty caused by the sky value. 
For simplicity we have assumed that the galaxy's brightness is constant
along the slit, although the background is slightly fainter on the Northern
part, best visible for the R-frame in Fig. \ref{slit}.   
Thus we underestimated the absorption by a tiny  amount. Since the gradients
in B and R are correlated, the reddening value moreover is only differentially
affected. Also   
the non-standard photometry is a second order effect because of our differential 
approach. The systematic uncertainty is thus smaller than the statistical one. 

\section{Discussion}

The above uncertainties support the claim that the reddening has a somewhat 
stronger wavelength dependence than the standard reddening law, in contrast to
the expectation of a greyer reddening law, if the dust patch was located in NGC 3269. A stronger wavelength
dependence points to a smaller grain size than in any ``standard'' extinction case.
In Bok globules one finds $R_V$-values as high as 6.5 \citep{strafella01},
compared to the standard value
of 3.1. This indicates that our cloud probably is not simply a distant Bok globule (which
then would be located in the Galactic halo).

On the other hand, although HI has not been detected in NGC 3269, the current upper limits of the HI column density reported
by \cite{barnes01} still do not rule out a considerable HI content of the dust
patch and its nature as a large complex in NGC 3269. They give their 1$\sigma$
 sensitivity as
$\rm 4.3 \cdot 10^{18} N_{HI} cm^{-2}$. Adopting  a 5$\sigma$ detection limit of
$\rm 2 \cdot 10^{19} N_{HI} cm^-2$ in a
87"x73" beam, the corresponding upper limit for the dust patch, normalizing
to area, is
$\rm 10^{22} N_{HI} cm^-2$. We adopt for
the ratio of V-absorption to total neutral hydrogen
column density $ A_V/N_H = 5.3 \cdot 10^{-22} cm^{-2}$ \citep{bohlin78,weingartner01}. 
With this value, the expected column density is $\rm 4 \cdot 10^{20} N_{HI}
 cm^{-2}$, which is anyway below the detection limit.

To our knowledge this would be the first detection of an isolated dust cloud
of this tiny angular size. The smallest globules in the surveys
of \cite{clemens88} and \cite{bourke95}
 have  diameters of about 30$\arcsec$.  
Larger projected dust structures have been seen before, for example against
Maffei 1 \citep{buta03}. Our case could be a rare coincidence.
On the other hand, structures of this size are
extremely difficult to see and the projection  onto an extended background
 source is practically the only
way to detect such objects. This is probably hopeless with spiral galaxies with a lot
of internal dust. Early type  galaxies at low galactic latitudes are
on the other hand not the most favoured objects in galaxy studies, so tiny
clouds may systematically be
overlooked. 

Our findings might gain interest in the context of studies of small molecular clouds.
\cite{heithausen04} recently presented interferometric observations
of ``small area molecular structures'' \citep{heithausen02}. He could resolve some tiny
molecular clouds with properties suggesting that this kind of molecular
structure is a common feature of the interstellar medium. The angular sizes
resemble that our dust patch, which, if in the Galaxy, is expected also to
be associated with molecular gas.  It is therefore possible that our cloud is
an optical counterpart to the molecular clumpuscules of Heithausen. According to him, 
``it is unknown, what creates or maintains these structures''. 

To get an idea of the mass associated
with the absorption in a ficticious distance, we adopt for
the ratio of V-absorption to total neutral hydrogen 
column density $ A_V/N_H = 5.3 \cdot 10^{-22} cm^{-2}$ \citep{bohlin78,weingartner01}. This holds for the diffuse ISM,
but the normalization should not drastically differ from that in dense clouds
\citep{weingartner01}. Thus, the total number of hydrogen
atoms is $N_{tot} = 1.9 \cdot 10^{21} \cdot r^2 \cdot \alpha^2 \cdot \pi \cdot A_V$, r being
the distance and $\alpha$ the angular radius. The mean R-absorption within our
slit is 0.2, value which we also use for the V-absorption.
An adopted distance of 100 pc and
a radius of 2" (corresponding to 200 AU) then suggest a mass of $\approx 10^{-5} M_{\odot}$.  
Given the nature of this estimate and the unknown distance, this mass is
not too different from the masses of the molecular clumps given by
Heithausen. The corresponding density is very high, about $ 0.5 \cdot 10^6 cm^{-3}$. 
\cite{heithausen02} noted that CO molecules in these clouds are not
likely to survive the interstellar radiation field unless there is sufficient
shielding. Our dust cloud could provide the necessary shielding.  
If the cloud is in the Galactic disk its distance can be as large as 1000 pc.
Its radius then would be 2000 AU, still significantly smaller than typical Bok
globules \cite[e.g.][]{strafella01}. 

\begin{figure}
\centering
\includegraphics[width=12cm,angle=00]{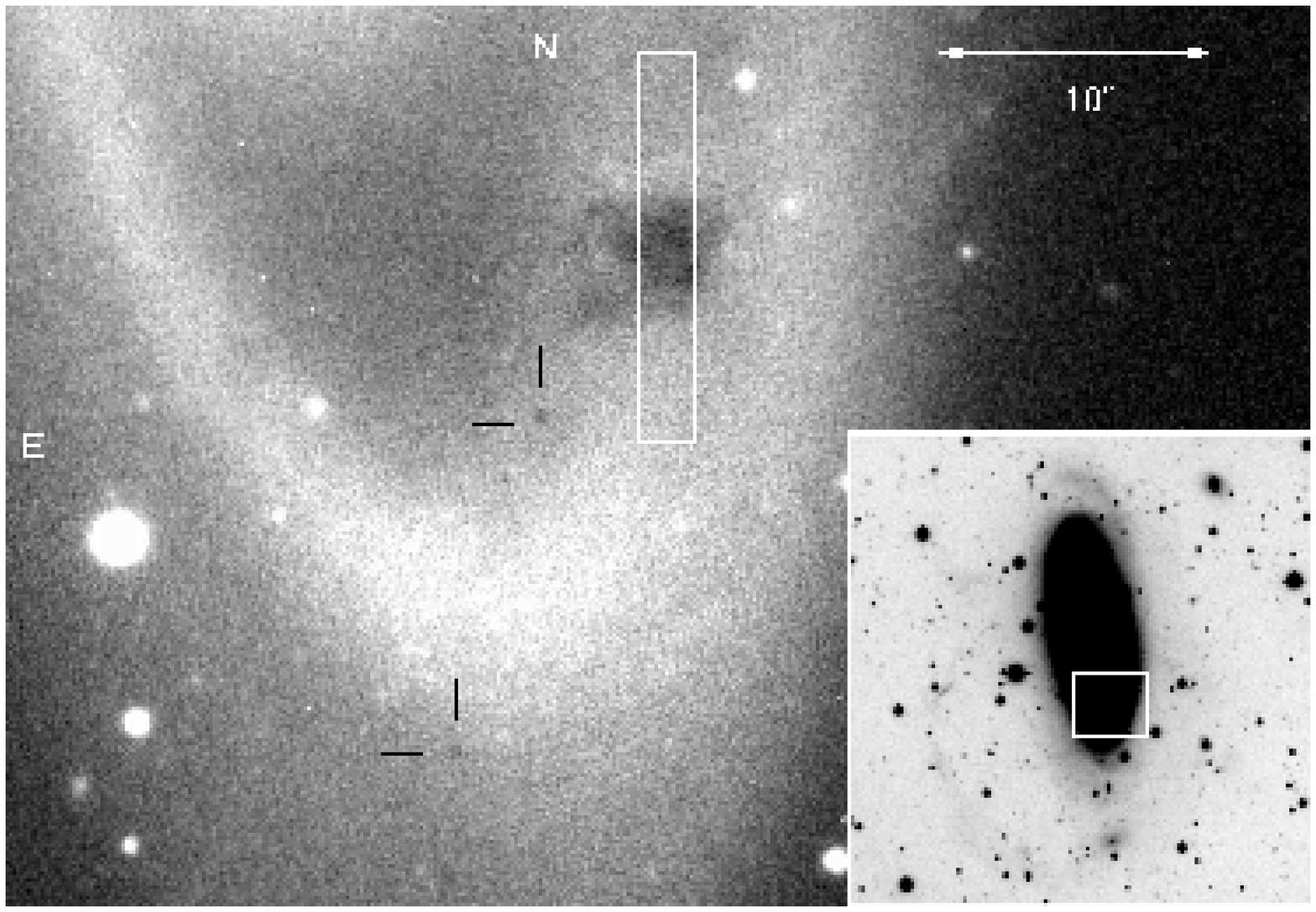}
\caption{The Southernmost  part of NGC 3269 showing the dust patch. Marked are
two even more tiny patches with sizes about 0.5$\arcsec$, which might be
associated with the larger feature. The slit along which we measure the 
absorption is indicated. The inset shows NGC 3269 on a larger scale in 
Washington C. Visible are various extended structures, particularly a 
large arc-like structure in the South-East. This morphology indicates 
previous or current tidal interactions so gas removal out of NGC 3269 is 
plausible. 
}

\label{image}
\end{figure}

The tiny clumpuscules indicated in Fig.~\ref{image}, if associated with the larger cloud, would have diameters
of the order 50 AU at 100 pc distance. They are too small for reliable absorption
measurements.

Whatever the nature of our dust structure may be, it would be interesting to
 search for more examples. Another candidate may be hosted by  NGC 3923,
hardly visible on images found in the NED, but prominently in recent
HST images (Th. Puzia, private communication). 

A related issue is perhaps the recent detection of small scale HI clouds 
\citep{braun05} in the local interstellar medium with extremely low column
 densities. The connection of this  new atomic feature to molecular or dust
structures is unknown. However, it shows that there are more small scale   
phenomena in the interstellar medium than previously thought.

\acknowledgments{
This research has made use of the NASA/ IPAC Infrared Science Archive, which is operated by the
 Jet Propulsion Laboratory, California Institute of Technology, under contract with the National
 Aeronautics and Space Administration.}


\clearpage

\clearpage

\begin{thebibliography}{}
\bibitem[Barnes \& Webster(2001)]{barnes01} Barnes, D. G., Webster, R. L. 2001, MNRAS, 324, 859
\bibitem[Bohlin et~al.(1978)]{bohlin78} Bohlin, R. C., Savage, B. D., Drake, J. F. 1978, ApJ 224, 132
\bibitem[Bourke et~al.(1995)]{bourke95} Bourke T.L., Hyland A.R., Robinson G. 1995, MNRAS 276, 1052
\bibitem[Braun \& Kanekar(2005)]{braun05} Braun R., Kanekar N. 2005, A\&A, in press (astro-ph/0505055)
\bibitem[Buta \& McCall(2003)]{buta03} Buta R., McCall M.L. 2003, AJ 125, 1150
\bibitem[Clemens \& Barvainis(1988)]{clemens88} Clemens D.P., Barvainis R. 1988, ApJS 68, 2570
\bibitem[D\'esert et al.(1988)]{desert88} D\'esert F.X., Bazell D., Boulanger F. 1988, ApJ 334, 815
\bibitem[Dirsch et~al.(2003)]{dirsch03} Dirsch B., Richtler T., Bassino L. 2003, A\&A 408, 929 
\bibitem[Dutra \& Bica(2002)]{dutra02} Dutra C.M., Bica E. 2002, A\&A 383, 631
\bibitem[He et~al.(1995)]{he95} He L., Whittet D.C.B., Kilkenny D., Spencer Jones J.H. 1995,
ApJS 101, 335
\bibitem[Heithausen(2002)]{heithausen02} Heithausen A., 2002, A\&A 393, L41
\bibitem[Heithausen(2004)]{heithausen04} Heithausen A., 2004, ApJ 606, L13
\bibitem[Rieke \& Lebofsky(1985)]{rieke85} Rieke G.H., Lebofsky M.J. 1985, ApJ 288, 618
\bibitem[Strafella et~al.(2001)]{strafella01} Strafella F., Campeggio L., Aiello S.,  Cecchi-Pestellini C. 2001,
ApJ 558, 717
\bibitem[Weingartner \& Draine(2001)]{weingartner01} Weingartner J.C., Draine B.T. 2001, ApJ 548, 296
\end{thebibliography}
\end{document}